\documentclass[final,5p,times,twocolumn,authoryear]{elsarticle}
\usepackage{hyperref}
\usepackage{pdfpages}
\usepackage{amssymb}
\usepackage{empheq}
\usepackage{lipsum}
\usepackage{float}
\usepackage{subcaption}
\usepackage{amsmath}
\setcitestyle{numbers}
\journal{arxiv}

\begin{document}

\begin{frontmatter}

\title{Chaotic studies in Magnetic Dipoles}

\author[first]{Reyhan Mehta}
\author[second]{Nishchal Dwivedi}
\affiliation[first]{{Department of Mechanical Engineering, SVKM’s NMIMS Mukesh Patel School of Technology Management and Engineering, Mumbai, India }}
\affiliation[second]{{Department of Basic Science and Humanities, SVKM’s NMIMS Mukesh Patel School of Technology Management and Engineering, Mumbai, India }}
\ead{nishchal.dwivedi@nmims.edu}

\begin{abstract}
The present work investigates the effect of an external rotating magnetic field on a magnetic needle, and aims to study chaotic behaviour. The equation of motion is modified to include damping and gravity. The bifurcation behaviour of such evolving systems is also studied. Furthermore, the coupling in pairs of magnetic needles is also analyzed. In addition, a 2-dimensional array of coupled needles is set up, and the synchronization properties of the population is studied. The study is completed by carrying out an investigation of the effects of noise on the synchronization of such a system.
\end{abstract}


\end{frontmatter}

\section{Introduction}
The phenomenon of magnetism has captivated mankind since it was first discovered in lodestones. Since then, mankind has been able to harness magnetism for several applications, ranging from navigation to modern day MRI machines. As we understand it today, magnetism essentially emerges from the spin and orbital motion of electrons in atoms, and thus a fully correct description of magnetism can only be supplied by the theory of quantum mechanics \cite{griffiths_schroeter_2018}.

When a single magnetic dipole is suspended in an external magnetic field, its resulting equation of motion is nonlinear\cite{griffiths}:
\begin{align}
    \mathbf{N} &= \mathbf{m} \times \mathbf{B} \\
    N_z = I\ddot \phi &= mB\sin \phi,
    \label{griff}
\end{align}
where, $\mathbf{N}$ is the torque acting on the dipole, $\mathbf{m}$ is the magnetic dipole moment, $\phi$ is the dipole's angular position, and $\mathbf{B}$ is the external magnetic field. Such simple systems have been well analyzed in the past \cite{griffiths}, \cite{Briggs}, \cite{Sang-Yoon}. 

In this work, the nonlinear behaviour of magnetic needles has been studied, with special emphasis on the combination of damping and gravitation, along with an analysis of the synchrony between such coupled dipoles. Nonlinear systems, in general, are of particular interest in understanding chaotic behaviour\cite{steven}, electrical circuits\cite{circuits},  synchronisation\cite{fireflies} and designing sensors and quantum devices \cite{sensors}. 

For this study, a magnetic needle is suspended in an external rotating magnetic field. To begin with, the equation of motion of the needle is written. Next, damping and gravity terms are introduced, and the resonance and bifurcation properties of the system are explored. The synchronization properties of this system have also been analyzed. Synchronization is when a set of oscillators which can interact with each other adjust their phases in such a way that their phase difference becomes constant over time \cite{steven}. Each oscillator can start with a random initial condition and independent phase. Given proper interaction strengths, it is observed that synchrony gradually appears with time. Synchronization can be seen in noisy circuits\cite{chua1993universal}, neuroscience \cite{phogat2018provoking} and fireflies \cite{craw} to name a few.

\section{Introduction to the system}
Consider a magnetic needle which is placed in a magnetic field perpendicular to its axis. The magnetic field rotates at a fixed angular frequency, and the needle itself is free to rotate. The equation of motion describing the needle is a second-order ordinary differential equation, \textit{with time dependence} (a nonautonomous equation) \cite{Wolfgang}:
\begin{equation}
\frac{d^2 \phi}{d t^2} + \frac{\mu}{I} B_0 \cos \omega t \sin \phi = 0,
\label{briggseqn}
\end{equation}
where $\phi$ is the angle of the needle in the field, $\mu$ its magnetic moment, $I$ its moment of inertia, and $B_0$ and $\omega$ are the amplitude and angular frequency of the external magnetic field, respectively. This equation results directly from an application of equation (\ref{griff}), by setting the angle of the rotating field to $\omega t$ and taking only the component of the field parallel to the needle's axis. Physical experiments have been carried out to investigate the range of values of these parameters which induce chaos in the needle's motion, by Briggs and others \cite{Briggs}, \cite{Sang-Yoon}.

It is convenient to define a parameter $\lambda \equiv \sqrt{2B_0\mu/(I\omega^2)}$, and then rewrite the equation of motion as
\begin{equation}
\frac{d^2 \phi}{d t^2} + \frac{\lambda^2 \omega^2}{2} \cos \omega t \sin \phi = 0.
\label{briggs2}
\end{equation}
The purpose of doing this is that an analytical result \cite{chirikov}, \cite{goldstein} shows that the motion of the needle in this simple case is chaotic precisely when $\lambda > 1.$ It is possible to rewrite this equation of motion in the form of three coupled first order equations, in the following way:
\begin{align}
\begin{split}
    x_0 &\equiv \phi\\
    x_1 &\equiv \dot \phi\\
    x_2 &\equiv t;
\end{split}
\end{align}
\textbf{The system of coupled differential equations:}

\begin{align}
\begin{split}
\dot x_0 &\equiv x_1\\
\dot x_1 &\equiv -\frac{1}{2}\lambda^2\omega^2\cos{(\omega x_2)}\sin(x_0)\\
\dot x_2 &\equiv 1.
\end{split}
\label{coupledeq}
\end{align}
\newline Having defined the three coupled variables $x_0$, $x_1$ and $x_2$, the system (\ref{coupledeq}) can be solved numerically as a system of differential equations. The results of this are shown in Figure 1.
\begin{figure}[h!]
    \centering
    \begin{subfigure}{0.4\textwidth}
        \centering
        \includegraphics[width=\textwidth]{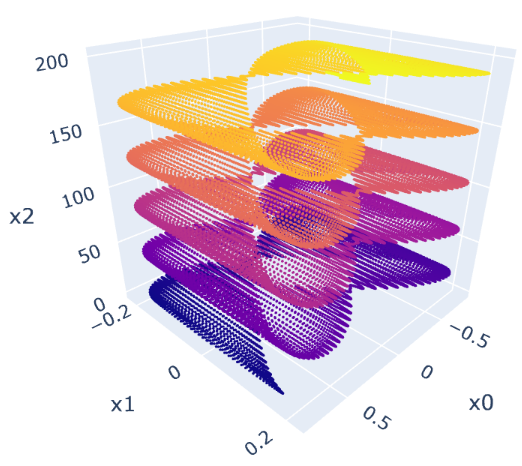}
        \caption{The phase space for $\lambda = 0.2$}
    \end{subfigure}
    \begin{subfigure}{0.4\textwidth}
        \centering
        \includegraphics[width=\textwidth]{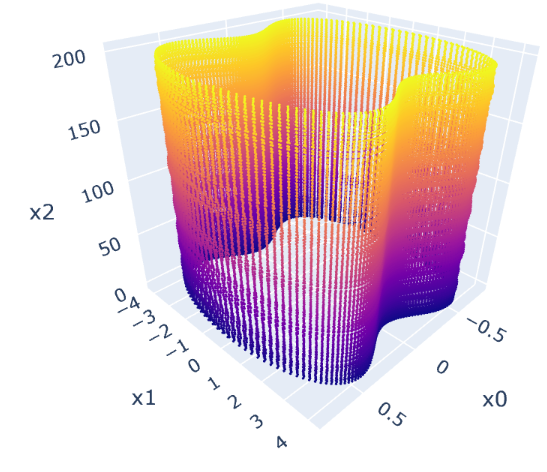}
        \caption{The phase space for $\lambda = \lambda_c = 1$}
    \end{subfigure}
    \caption{Numerical solution of the coupled system of differential equations, for $\omega = 4\pi$. For $\lambda < 1$, we get multiple periodic closed loops in the phase space, corresponding the non-chaotic behaviour; for $\lambda_c = 1$, we get a single loop in the phase space, which corresponds to the critical point where bifurcation occurs. Beyond $\lambda > 1$, the motion of the needle is chaotic.}
\end{figure}
\section{Simulation of the Briggs Experiment}
In order to determine the conditions under which a system demonstrates chaos, the typical method is to plot a \textit{Poincar\'e Map}, which is a specialized phase space, where only those points are plotted which correspond to one complete revolution of the external driving agency - in this case, the external magnetic field. This means that points are plotted only at times $t = 2\pi n/\omega$ \cite{gerald}.
If the motion of the system is periodic, a set of points will appear on the Poincar\'e Map in accordance with a periodic pattern. On the other hand, if the motion is chaotic, the points on the map will look like a random diffuse scatter over all space.

Equation (\ref{briggs2}) is a nonlinear, second-order ordinary differential equation, where the second derivative depends on $\phi$ and $t$ (in the more general case discussed below, $\ddot \phi$ depends on $\dot \phi, \phi,$ and $t$). The system described here is highly sensitive to initial conditions, and the equation of motion cannot be solved analytically, so these equations are solved numerically \cite{Wolfgang}, \cite{dan}.

\begin{figure}[h!]
    \centering
    \begin{subfigure}{0.4\textwidth}
        \centering
        \includegraphics[width=\textwidth]{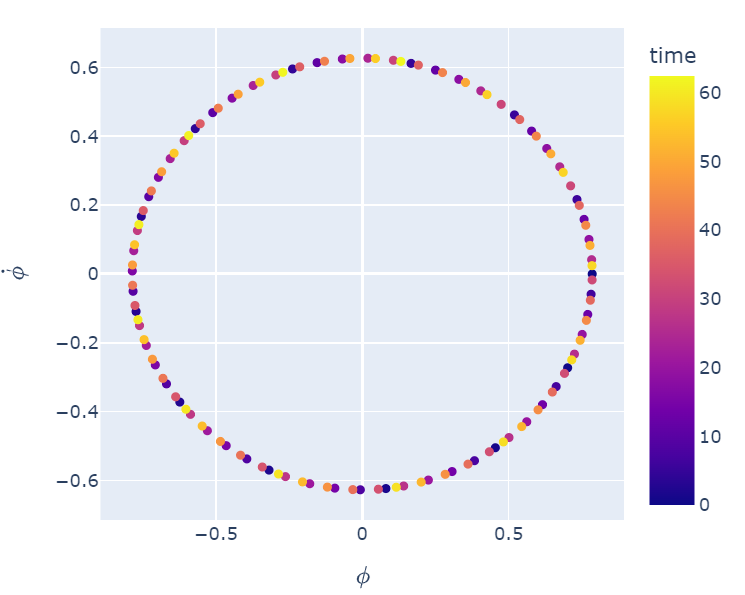}
        \caption{The Poincar\'e map for $\lambda = 0.5$}
    \end{subfigure}
    \begin{subfigure}{0.4\textwidth}
        \centering
        \includegraphics[width=\textwidth]{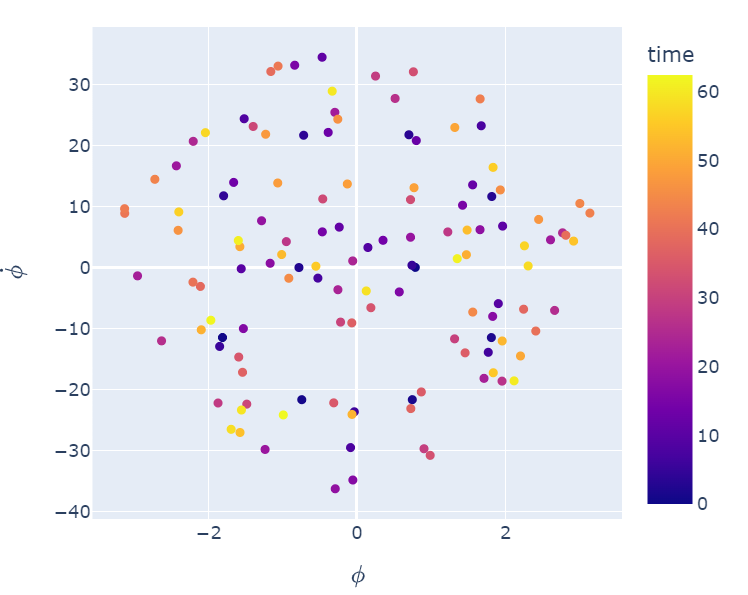}
        \caption{The Poincar\'e map for $\lambda = \sqrt{3}$}
    \end{subfigure}
    \caption{Simulation of the simple magnetic needle. The Poincar\'e map for $\lambda = 0.5 < 1$ is a ring, corresponding to periodic behaviour of the dipole. For $\lambda = \sqrt{3} > 1$, the Poincar\'e map is a diffuse scatter of points, corresponding to non-periodic, chaotic motion of the dipole.}
    \label{poincare1}
\end{figure}

The simulation was performed for the initial conditions $(\phi_0, \dot \phi_0) = (\pi/4, 0)$, for $\omega = 4\pi$. The results are in complete agreement with the theory and physical experiments carried out earlier; they are shown in Fig. \ref{poincare1}. In Fig. 2(a), the Poincar\'e map shows periodicity where the points arrange themselves in an oraganised manner. Fig. 2(b) shows no such periodicity, hence indicating chaos.

\section{Inclusion of Damping}
A more realistic picture of such a system will comprise of damping. To study this system in more detail, we include a damping term, $\gamma \dot \phi$, where $\gamma$ is the damping strength, in the equation of motion. 
\begin{equation}
    \frac{d^2 \phi}{d t^2} + \gamma \frac{d\phi}{dt} + \frac{\lambda^2 \omega^2}{2} \cos \omega t \sin \phi = 0.
\end{equation}
\begin{figure}[h!]
    \centering
    \begin{subfigure}{0.4\textwidth}
        \centering
        \includegraphics[width=\textwidth]{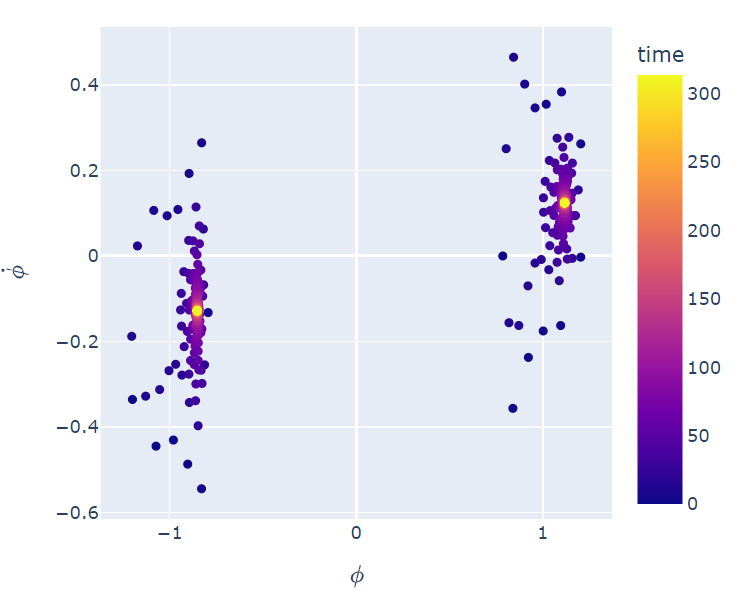}
        \caption{The Poincar\'e map for $\lambda = 2$}
    \end{subfigure}
    \begin{subfigure}{0.4\textwidth}
        \centering
        \includegraphics[width=\textwidth]{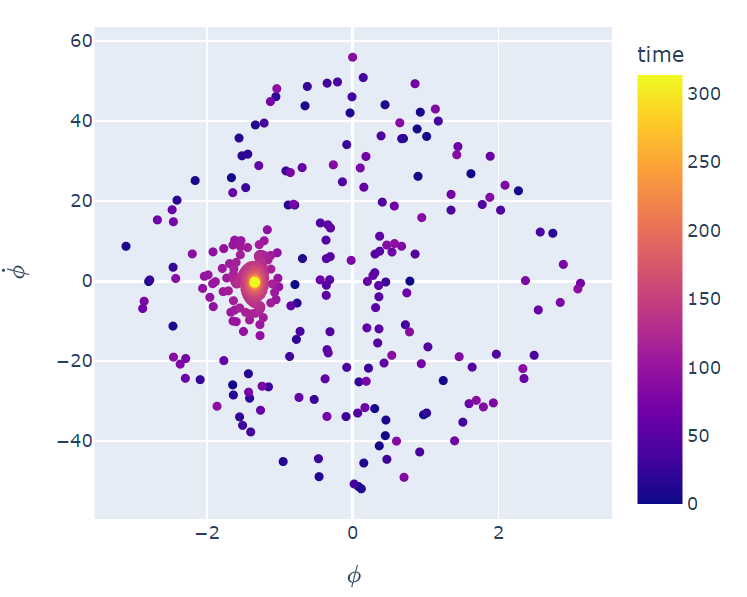}
        \caption{The Poincar\'e map for $\lambda = 3$}
    \end{subfigure}
    \begin{subfigure}{0.4\textwidth}
        \centering
        \includegraphics[width=\textwidth]{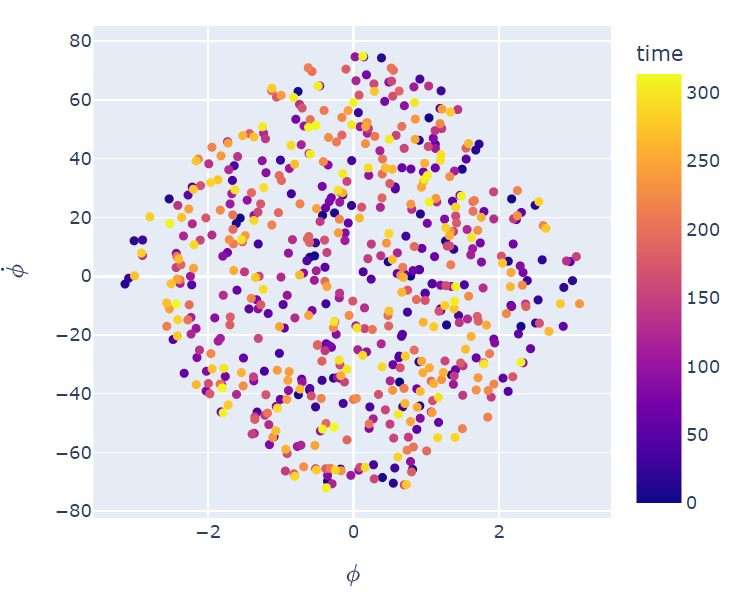}
        \caption{The Poincar\'e map for $\lambda = 4$}
    \end{subfigure}
    \caption{Simulation of the needle with damping ($\gamma = 0.05$). The dipole is able to retain periodic motion for values of $\lambda$ larger than 1, since the damping force dissipates energy from the driving magnetic field. In this case, $\lambda_c = 3.4$.}
    \label{poincare2}
\end{figure}
As is evident from the resulting Poincar\'e maps (see Fig. \ref{poincare2}), the following result is found. The magnetic needle \textit{resists} chaos even for $\lambda$ values greater than 1, which can be explained by supposing that the damping force dissipates energy from the driving magnetic field, so a larger amplitude is required to induce chaos. This is demonstrated in Fig. \ref{poincare2}. In this case, where $\gamma = 0.05$, the critical value of $\lambda$ for the onset of chaos was found to be $\lambda_c = 3.4$. It is also interesting to note that for $\lambda = 3$, there is perfect mode-locking between the needle and the magnetic field - it can be seen in the Poincar\'e map that after the transient motion has faded, the magnetic needle only appears at a single point in its trajectory, in the Poincar\'e map, which means that its frequency is exactly the same as the frequency of the driving magnetic field.

\section{Inclusion of the gravitational term}
The next step is to further generalise the equation of motion by adding an additional harmonic influence due to gravity. This term corresponds to the natural frequency of oscillation of the magnetic needle, and is analogous to the gravitational term $\omega_0^2 \sin \theta$ in the equation of motion of a freely swinging pendulum in a gravitational field, where $\omega_0$ is the natural frequency for small angle approximation.

\begin{equation}
    \frac{d^2 \phi}{d t^2} + \gamma \frac{d\phi}{dt} + \left(\frac{\lambda^2 \omega^2}{2} \cos \omega t + \omega_0^2 \right)\sin \phi = 0.
\end{equation}
Physically, this term can be introduced into the system by suspending the needle in a gravitational field in a way such that the axis of rotation of the needle is perpendicular to the gravity, so that the needle itself acts like a swinging pendulum under the influence of gravity.

\begin{figure}[h!]
    \centering
    \begin{subfigure}{0.4\textwidth}
        \centering
        \includegraphics[width=\textwidth]{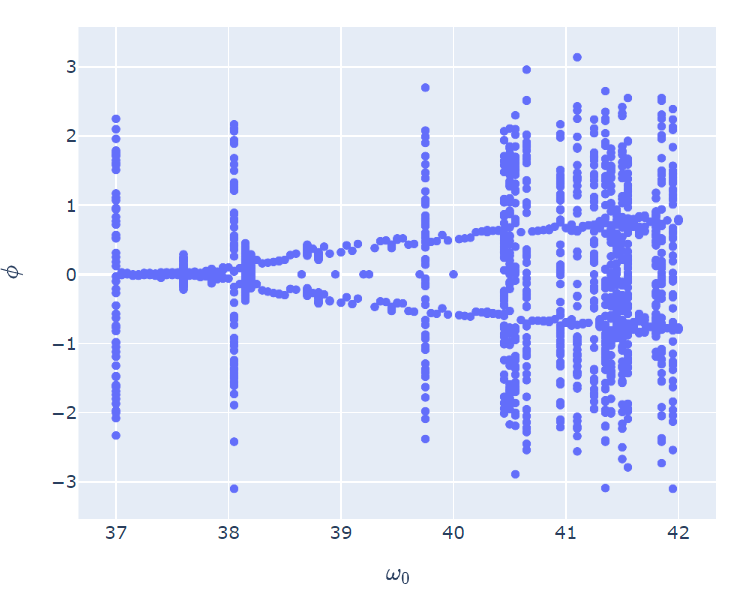}
    \end{subfigure}
    \caption{The bifurcation diagram of $\omega_0$ vs. $\phi$. A bifurcation takes place at $\omega_0 \approx 37.5$. In addition to the bifurcation, thin slivers of chaos are also superposed onto the bifurcation diagram. The system is highly sensitive to changes in $\omega_0$ near these thin slivers.}
    \label{bif1}
\end{figure}

In order to investigate the effect of $\omega_0$ on the periodicity of the needle's motion, we plot a \textit{bifurcation diagram} of $\phi$ vs. $\omega_0$. What this means is that the needle's motion is simulated for a long time, for a particular value of $\omega_0$, and then all of the occurring values of $\phi$ in the needle's resulting Poincar\'e map (after the initial transient points are discarded) are plotted in a vertical line for that particular value of $\omega_0$. The values of the other system parameters have been fixed to:
\begin{enumerate}
    \item $\lambda = 3$
    \item $\gamma = 0.05$
    \item $\omega = 8\pi$,
\end{enumerate}
so that the only independent parameter which is being varied is $\omega_0$, which ranges from $\omega_0 = 37$ to $\omega_0 = 42$, in steps of $\Delta \omega_0 = 0.05$. As before, the initial conditions remain to be fixed at $(\phi, \dot \phi) = (\pi/4, 0).$ The resulting bifurcation diagram is shown in Fig. \ref{bif1}.
It can be seen in this bifurcation diagram that a distinct period doubling takes place at $\omega_0 \approx 37.5$. This period 2 region persists throughout the range of values selected for $\omega_0$. However, it can also be seen that thin slivers corresponding to chaotic motion are superposed onto this period 2 region, so the needle seems to exhibit both period 2 motion and chaotic motion within this range of values of $\omega_0.$

\begin{figure}[h!]
    \centering
    \begin{subfigure}{0.4\textwidth}
        \centering
        \includegraphics[width=\textwidth]{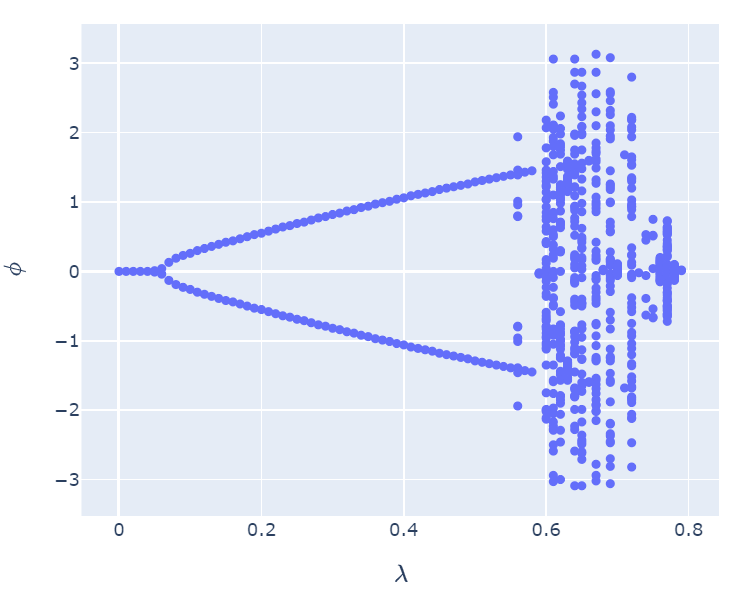}
    \end{subfigure}
    \caption{The bifurcation diagram of $\lambda$ vs. $\phi$. A period doubling bifurcation takes place at $\lambda = 0.06$, when the magnetic field becomes strong enough to move the needle against damping. At $\lambda = 0.6$, the system becomes chaotic and is highly sensitive to changes in $\lambda$.}
    \label{bif2}
\end{figure}

In addition to plotting the bifurcation diagram for $\omega_0$ vs. $\phi$, it was also decided to plot the bifurcation diagram for $\lambda$ vs. $\phi$, for the case where the harmonic term is present. The following plot (Fig. \ref{bif2}) shows the result. The other parameters were fixed as:
\begin{enumerate}
    \item $\omega=8\pi$
    \item $\omega_0 = 4\pi$
    \item $\gamma=0.05$
    \item $(\phi_0, \dot \phi_0) = (\pi/4, 0)$
\end{enumerate}
It can be seen that for values of $\lambda$ less than 0.06, the magnetic field is too weak to drive the needle against the gravity and damping; however, at $\lambda = 0.06$, a distinct period doubling bifurcation occurs and the needle is able to move periodically against the damping and gravity. At $\lambda \approx 0.6$, the motion of the needle becomes chaotic and its period tends to infinity.
\section{Coupled Magnetic Needles}
\subsection{Coupled motion of needle twins}
\begin{figure}[h!]
    \centering
    \begin{subfigure}{0.475\textwidth}
        \centering
        \includegraphics[width=\textwidth]{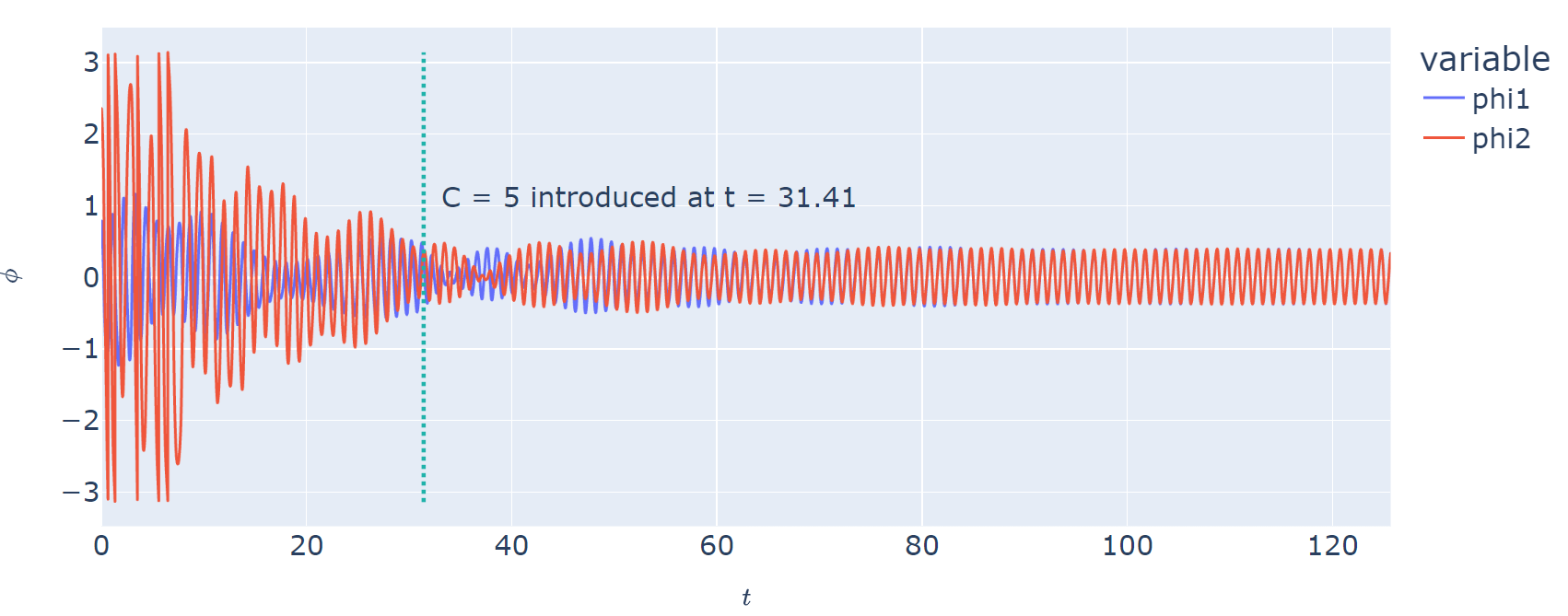}
    \end{subfigure}
    \begin{subfigure}{0.475\textwidth}
        \centering
        \includegraphics[width=\textwidth]{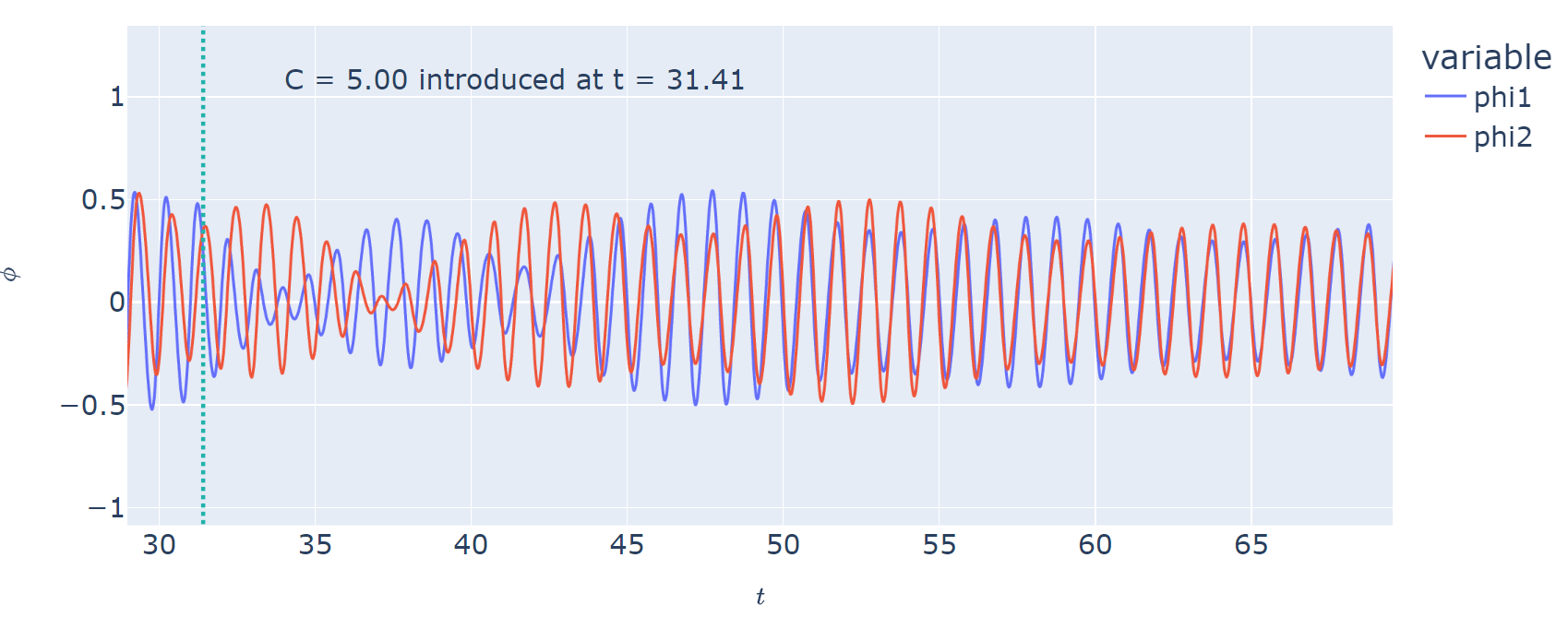}
    \end{subfigure}
    \caption{The trajectory for the coupled twin pair. $C=5$ is introduced at $t = 31.41$. It can be seen that the phase difference between these coupled dipoles asymptotically becomes zero - this is the first mode of synchronization.}
    \label{traj1}
\end{figure}

Having completely analyzed the motion of a single magnetic needle, suspended in a magnetic and gravitational field, with damping, we can now proceed to analyze the motion of two such magnetic needles which are coupled together. The primary interest of this part of the study was to investigate which conditions are able to induce synchrony in the twin pair of magnetic needles.
To begin with, the equation of motion of each needle is modified, by adding an extra anti-symmetric \textit{coupling term} to both equations. The resulting equation of motion for both dipoles looks like
\begin{align*}
\frac{d^2 \phi_1}{d t^2} + \gamma \frac{d \phi_1}{dt} + \left(\omega_{0}^2 + \frac{\lambda^2 \omega^2}{2} \cos \omega t\right) \sin \phi_1 + C\sin(\phi_1 - \phi_2) &= 0,\\
\frac{d^2 \phi_2}{d t^2} + \gamma \frac{d \phi_2}{dt} + \left(\omega_{0}^2 + \frac{\lambda^2 \omega^2}{2} \cos \omega t\right) \sin \phi_2 + C\sin(\phi_2 - \phi_1) &= 0,
\end{align*}
where $C$ is a positive constant, known as the \textit{coupling constant.} The chosen form of the coupling is sinusoidal, because it ensures that the interaction torque between the two magnetic needles is anti-symmetric, in accordance with the conservation of angular momentum.
Following this, a program is written for the needle twins and a numerical simulation is carried out with the following initial conditions.
\begin{enumerate}
  \item $\lambda = 1$
  \item $\omega = 2\pi$
  \item $\omega_0 = 2\pi$
  \item $\gamma = 0.10$
  \item $C = 5$
  \item $\phi_1(0) = \pi/4$
  \item $\phi_2(0) = 3\pi/4$
  \item $\dot\phi_1(0) = \dot\phi_2(0) = 0.$
\end{enumerate}
The resulting trajectory of the needles' motion is depicted in Fig. \ref{traj1}.
For the purposes of this study, synchronization is defined as a state achieved by a set of coupled oscillators when the difference between their phases becomes constant asymptotically \cite{synch}.
The coupling term is introduced into the system at $t=10\pi$. It is evident from the Fig \ref{traj1} that the coupling between the needles results in their motion getting synchronized.
\subsection{Modes of Synchronization}
\begin{figure}[h!]
    \centering
    \begin{subfigure}{0.475\textwidth}
        \centering
        \includegraphics[width=\textwidth]{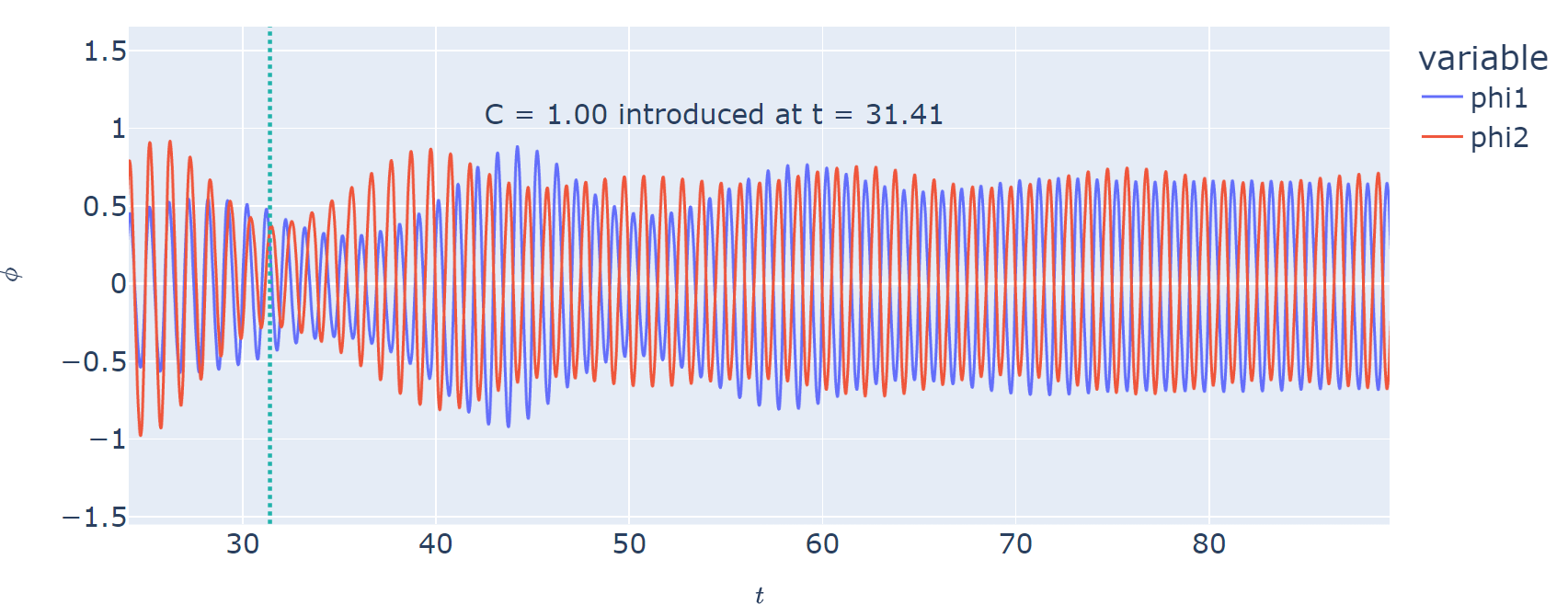}
    \end{subfigure}
    \begin{subfigure}{0.475\textwidth}
        \centering
        \includegraphics[width=\textwidth]{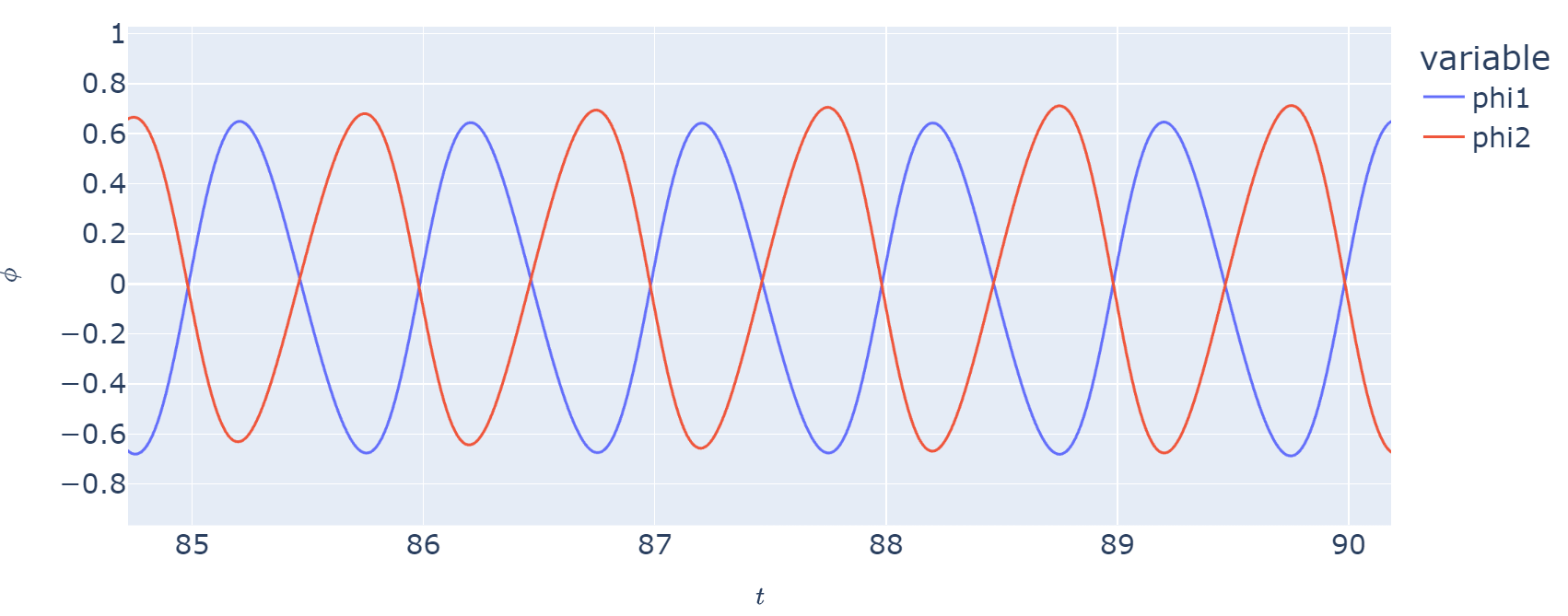}
    \end{subfigure}
    \caption{The trajectory for the coupled twin pair, with a different coupling constant. $C=1$ is introduced at $t = 31.41$. It can be seen that the phase difference between these coupled dipoles asymptotically becomes $\pi$ radians - this is the second mode of synchronization.}
    \label{traj2}
\end{figure}

While the analysis for the previous section was being carried out, it was also discovered that the magnetic needle in question demonstrates various \textit{modes of synchronization}. What this means is that the limiting phase difference between the needles appears to take on different values, depending on the nature of the coupling constant. For instance, when a coupling constant of $C=1$ is taken, the limiting phase difference between the needles is not $0$ (as in Fig. \ref{traj1}), but rather, it is half a cycle (as shown in Fig. \ref{traj2}). These different modes of synchronization have analogues in the quantum theory of magnetism - the first mode is akin to ``ferromagnetism,'' and the second one resembles ``antiferromagnetism'' \cite{nature}.
\subsection{Stability analysis of coupled dipoles}
The energy of a magnetic dipole in a magnetic field $\mathbf{B}$ is
\begin{equation}
    U = -\mathbf{m}\cdot\mathbf{B},
\end{equation}
where $\mathbf{m}$ is the magnetic dipole moment, $\mathbf{B}$ is the external magnetic field, and $U$ is the interaction energy \cite{griffiths}. From this expression, the interaction energy energy of two magnetic dipoles separated by a displacement $\mathbf{r}$ is given by 
\begin{equation}
    U = \frac{\mu_0}{4\pi r^3}\left[\mathbf{m_1}\cdot\mathbf{m_2} - 3(\mathbf{m_1}\cdot\mathbf{\hat r})(\mathbf{m_2}\cdot\mathbf{\hat r}) \right].
\end{equation}
This can be re-expressed in terms of the angles $\phi_1$ and $\phi_2$ the dipoles make with respect to their separation vector $\mathbf{r}$:
\begin{equation}
    U = \frac{\mu_0}{4\pi r^3}\left[\sin \phi_1 \sin \phi_2 - 2\cos \phi_1 \cos \phi_2 \right].
\end{equation}
This result can be used to find the stable configuration two dipoles would adopt if held a fixed distance apart, but left free to rotate. The equilibria of the system can be found by setting the gradient of the potential energy function to zero (since $\mathbf{F} = -\nabla U$):
\begin{align*}
    \nabla U = \frac{\partial U}{\partial \phi_1}\mathbf{\hat \phi_1} + \frac{\partial U}{\partial \phi_2}\mathbf{\hat \phi_2} &= \mathbf{0} \\
    \implies \sin(\phi_1) = \sin(\phi_2) &= 0 \\ 
    \text{or } \cos(\phi_1) = \cos(\phi_2) &= 0.
\end{align*}
In order to investigate the stability of these points, the formal procedure requires us to calculate the second order partial derivatives of the potential function \cite{riley_hobson_bence_2002}, \cite{boas2006mathematical}. However, we choose to instead plot the 3-dimensional surface $U(\phi_1, \phi_2)$ in $\phi_1, \phi_2$ space, to visually investigate which points are stable and which ones are unstable in an intuitive fashion.
\begin{figure}[h!]
    \centering
    \begin{subfigure}{0.4\textwidth}
        \centering
        \includegraphics[width=\textwidth]{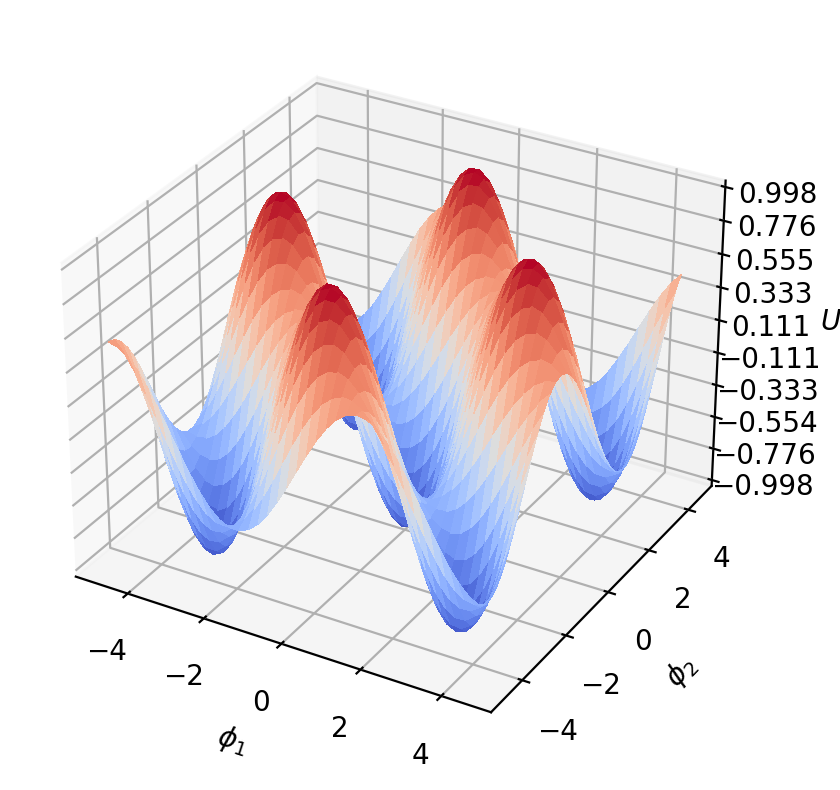}
    \end{subfigure}
    \caption{The 3D surface plot of $U(\phi_1, \phi_2)$. The valleys of the potential function (the points of stable equilibria) correspond to $\phi_1, \phi_2 = n\pi$ for integral $n$, which occurs when the needles are both parallel and aligned with their separation vector $\mathbf{r}.$}
    \label{3d}
\end{figure}
It is evident from Fig. \ref{3d} that $U$ attains a minimum whenever $\phi_1, \phi_2 = n\pi$ for integral $n$, and hence these are the points of stable equilibrium. Each of these points correspond to the needles aligned in a parallel fashion (along the separation vector), which explains why the first mode of synchronization is stable in the case of coupled needle twins.

\subsection{2-dimensional array of coupled dipoles}
\subsubsection*{Overview}
We now ask the question, how will a 2D array of such magnets behave and can it exhibit synchronisation. A 2-dimensional array of magnetic needles is considered, where every needle is coupled to its immediate neighbours. We make a circular symmetry by coupling every needle on the edges of the array to the needle immediately on the directly opposite edge, so that the array is topologically equivalent to a \textit{torus} \cite{kibble2004classical}. Topology can play a crucial role in the synchronization of a population of coupled oscillators \cite{top}, and it is anticipated here that connecting the edges in this way will improve the synchronization process, by allowing every needle to have the same number of ``immediate neighbours.''
\subsubsection*{The equations of motion}
A total of four independent coupling terms will have to be added to the equation of motion of each needle (left, right, top, bottom); the equation of motion of the needle in the $(i,j)^{\text{th}}$ position looks like:
\begin{align*}
\ddot \phi_{i,j} + \gamma \dot \phi_{i,j} + \left(\omega_{0}^2 + \frac{\lambda^2 \omega^2}{2} \cos \omega t\right) \sin \phi_{i,j} + C_{i,j} &= 0, \\
\text{where } C_{i,j} = C[\sin(\phi_{i,j} - \phi_{i+1, j}) + \sin(\phi_{i,j} - \phi_{i-1, j}) &+ \\
\sin(\phi_{i,j} - \phi_{i, j+1}) + \sin(\phi_{i,j} - \phi_{i, j-1})] &= 0,
\end{align*}
where $0 \leq i \leq \text{height}$ and $0 \leq j \leq \text{width}$. It is understood in this context that negative indices correspond to wrapping around the array and emerging from the other side.
\subsubsection*{Carrying out the simulation}

To carry out the analysis, an $8\times 8$ array of coupled needles is initialized. Every needle is started with zero velocity, and its angle is chosen randomly between $0$ and $2\pi$. The resulting \textit{phasemap} which describes the state of this array at time $t = 0$ and at $t = 200$ is shown in the following figures; the phase of every needle is encoded by the colour of its corresponding square, with the phase-to-color map shown by the vertical colour scale. Incidentally, such phase maps are very useful in analyzing other phenomena where coupled oscillators play a central role in the evolution of a physical system, such as in cardiac fibrillation \cite{heart}. 

\subsubsection*{Resulting phasemap}
\begin{figure}[h!]
    \centering
    \begin{subfigure}{0.2\textwidth}
        \centering
        \includegraphics[width=\textwidth]{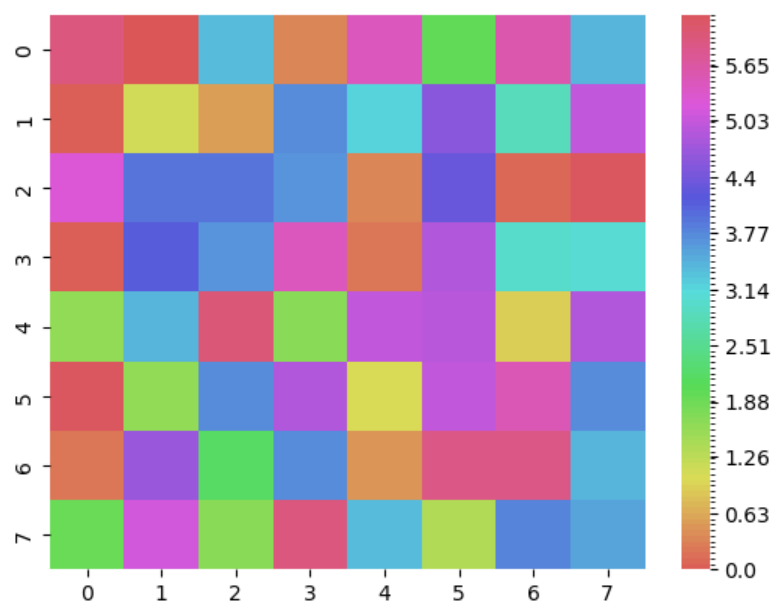}
        \caption{The phase map at $t = 0$}
    \end{subfigure}
    \begin{subfigure}{0.2\textwidth}
        \centering
        \includegraphics[width=\textwidth]{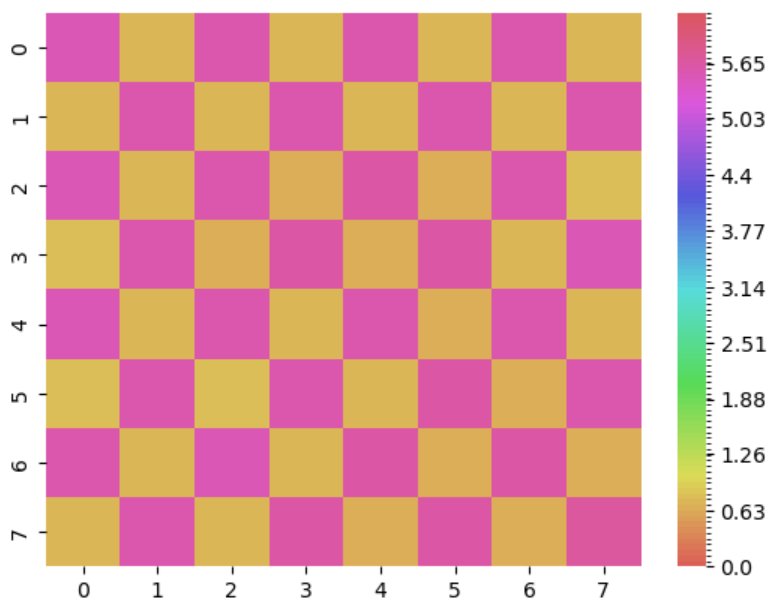}
        \caption{The phase map at $t = 200$}
    \end{subfigure}
    \begin{subfigure}{0.4\textwidth}
        \centering
        \includegraphics[width=\textwidth]{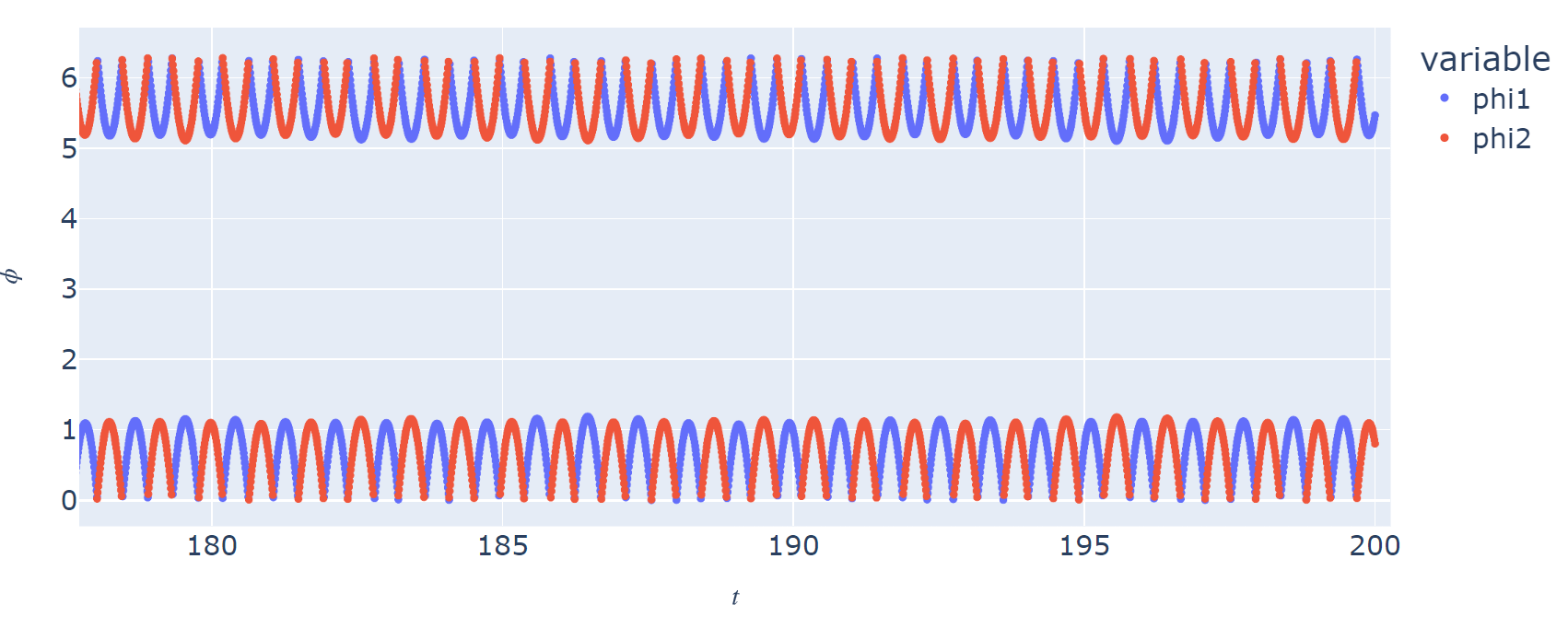}
        \caption{The trajectory for two neighbours, during the last 20 seconds.}
    \end{subfigure}
    \caption{The second mode of synchronization, as discovered in the dipole array. Over time, the dipoles are all moving exactly out of phase with each of their immediate neighbours.}
    \label{board1}
\end{figure}

Fig. \ref{board1} shows that every needle synchronizes to all of its immediate neighbours in exactly the same way as the twin needles synchronized to each other in the previous section. The phasemap shown in Fig. \ref{board1} corresponds to the \textit{second} mode of synchronization, where every needle is exactly half a cycle away from its neighbours.
These are the conditions which were required to induce the second mode of synchronization in the needle array:
\begin{enumerate}
    \item $C = 5$
    \item $\lambda = 0.5$
    \item $\omega = \omega_0 = 2\pi$
    \item $\gamma = 0.05$.
\end{enumerate}
The ``antiferromagnetic'' state observed for the needles is similar to the pattern of alignment of electron spins in a material which is antiferromagnetic, and it is suspected that the mechanism of electron alignment is similar to classical dipole alignment observed in this study \cite{beth}.
\subsection{Including noise in the magnetic field}
It is interesting to investigate the consequences of including \textit{noise} in the magnetic field. This part of the study was carried out because including noise in the field of a coupled oscillator system can have a significant influence on its dynamics\cite{noise}. In order to do this, the equation of motion of each needle must be slightly altered in the following way:
\begin{align*}
    \ddot \phi_{i,j} + \gamma \dot \phi_{i,j} + \left[\omega_{0}^2 + \frac{\lambda^2 \omega^2}{2} (\cos \omega t + N(t))\right] \sin \phi_{i,j} + C_{i,j} &= 0,
\end{align*}
where $N(t)$ represents the \textit{noise function}. The function chosen for this application was the Gaussian Distribution, with mean $\mu = 0$ and standard deviation $\sigma = 0.2$. What this means is that every time a single step is to be executed in the simulation, a value is chosen randomly (using a pseudo-random generator), and this value is then taken as $N(t)$ for that particular step.

\begin{figure}[h!]
    \centering
    \begin{subfigure}{0.475\textwidth}
        \centering
        \includegraphics[width=\textwidth]{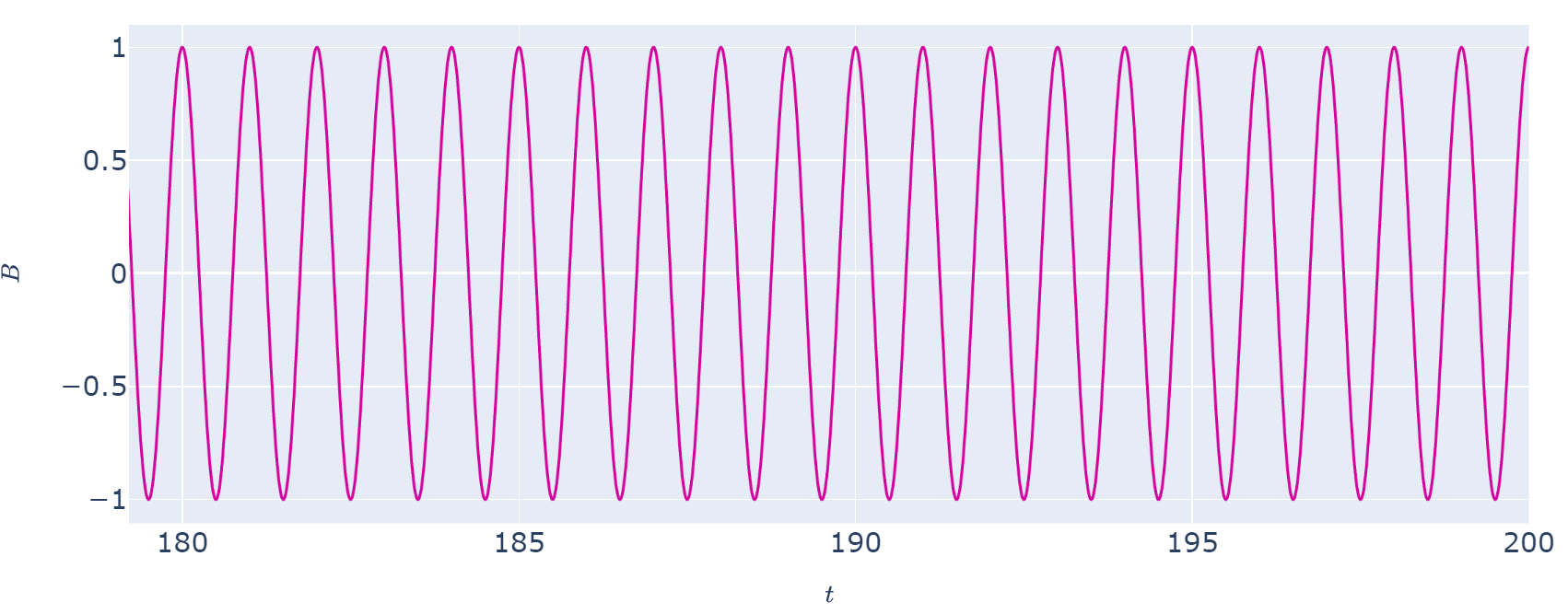}
        \caption{The input signal, before superposing noise to the periodic input signal.}
    \end{subfigure}
    \begin{subfigure}{0.475\textwidth}
        \centering
        \includegraphics[width=\textwidth]{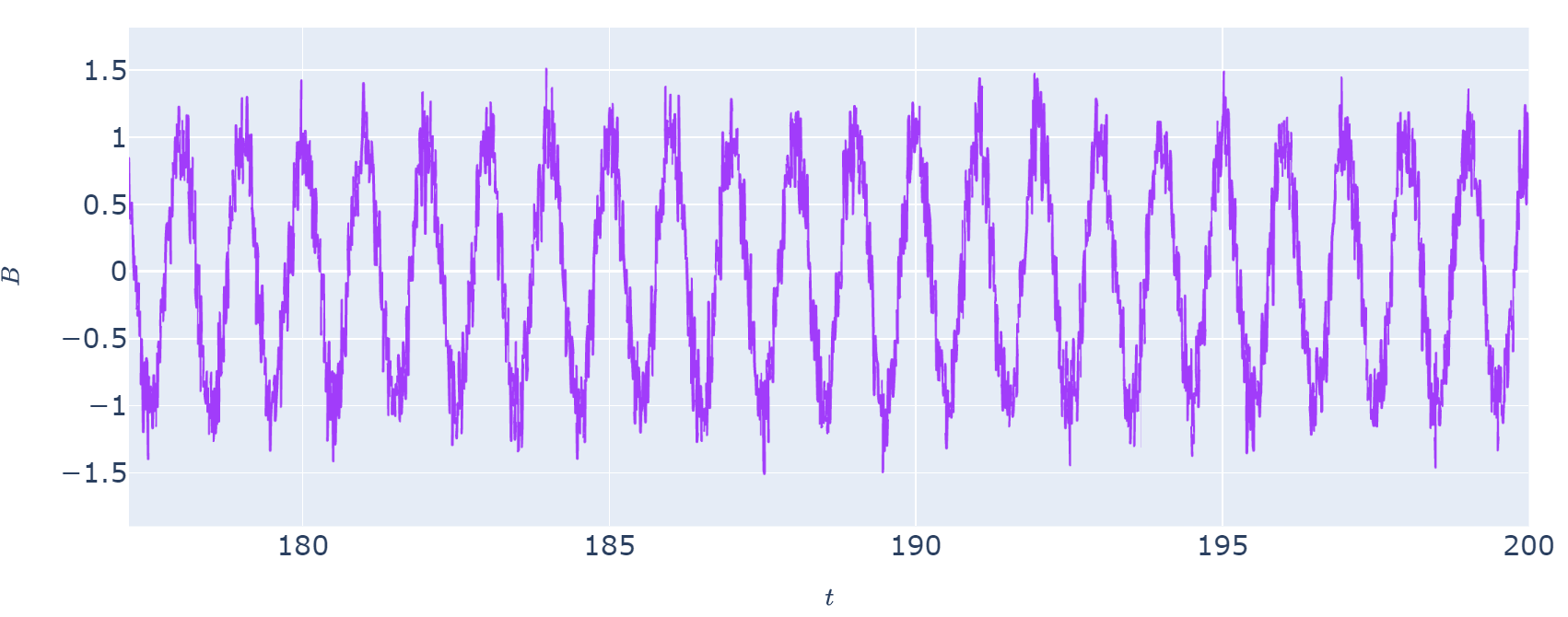}
        \caption{The resulting input signal, after superposing noise to the original input signal.}
    \end{subfigure}
    \caption{The magnetic field as a function of time.}
    \label{sig}
\end{figure}
\begin{figure}[h!]
    \centering
    \begin{subfigure}{0.2\textwidth}
        \centering
        \includegraphics[width=\textwidth]{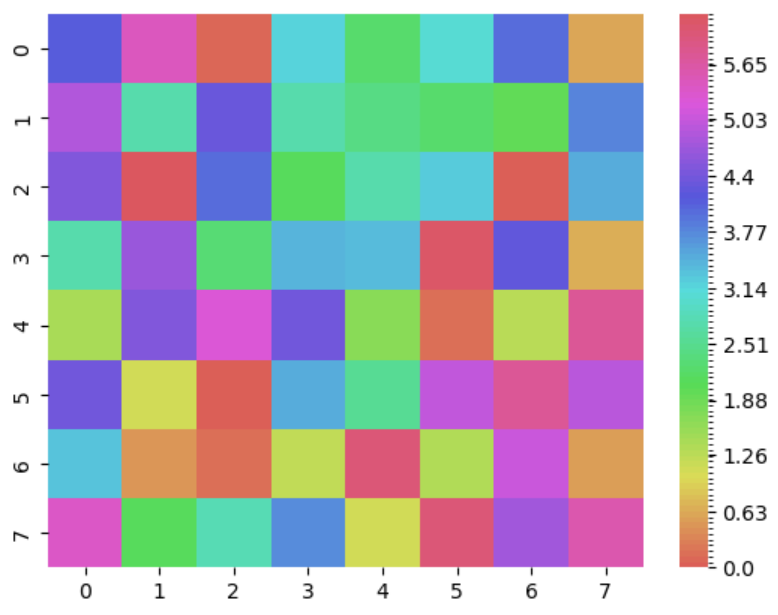}
        \caption{The phase map at $t = 0$}
    \end{subfigure}
    \begin{subfigure}{0.2\textwidth}
        \centering
        \includegraphics[width=\textwidth]{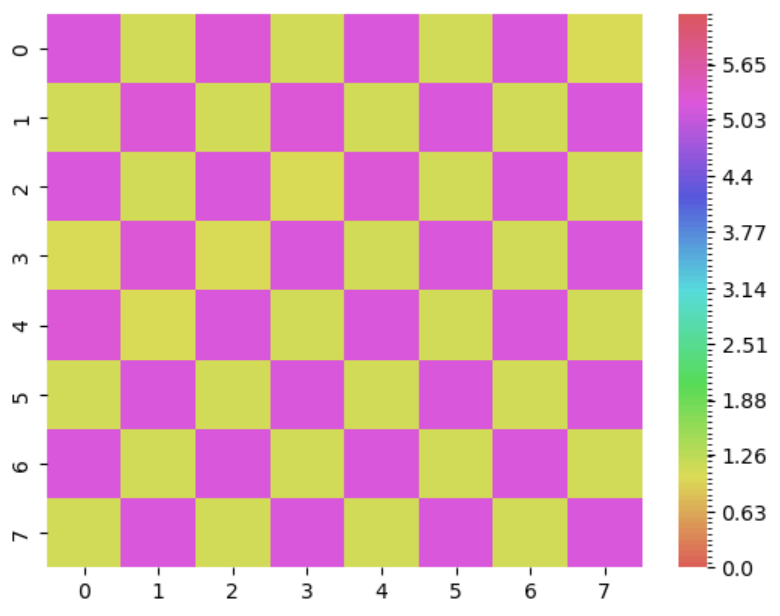}
        \caption{The phase map at $t = 200$}
    \end{subfigure}
    \begin{subfigure}{0.4\textwidth}
        \centering
        \includegraphics[width=\textwidth]{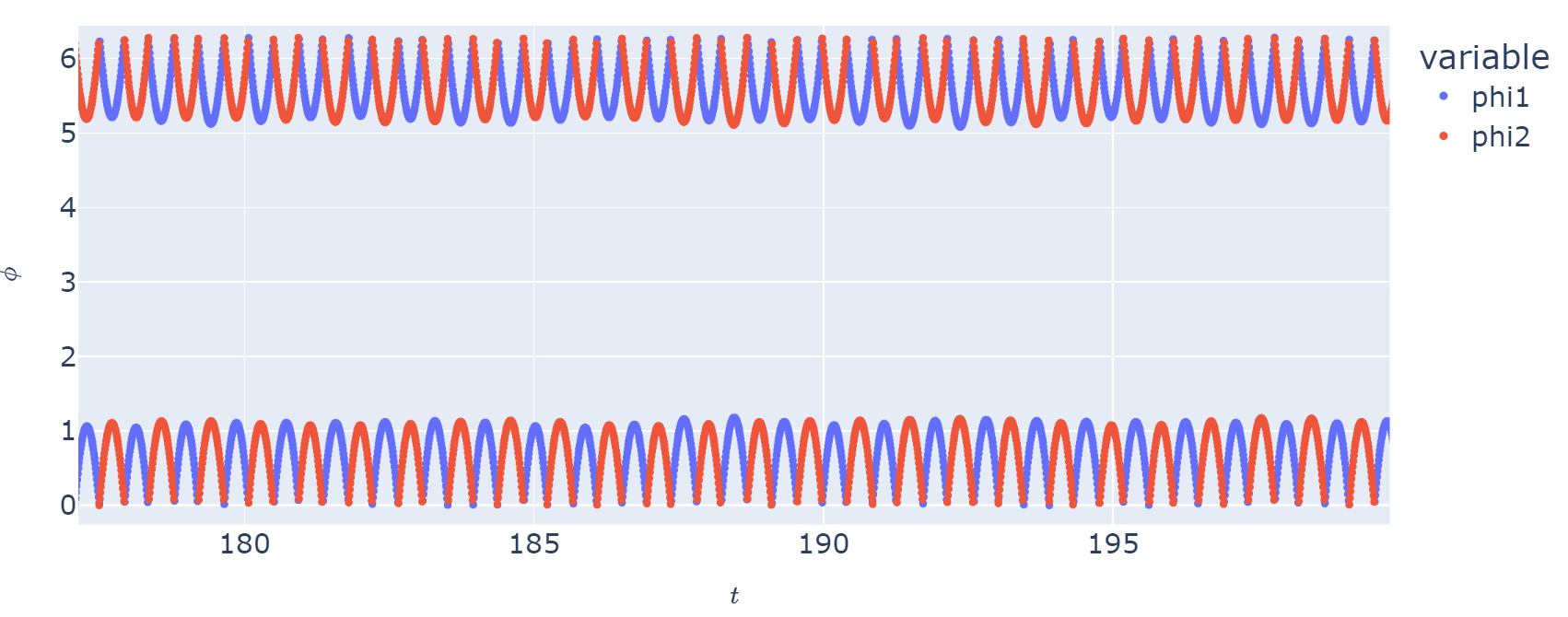}
        \caption{The trajectory for two neighbours, during the last 20 seconds.}
    \end{subfigure}
    \caption{The needle array, with noise included in the signal. Even after imposing Gaussian noise onto the magnetic field, the dipoles still attain the second mode of synchronization, as was observed in Fig. \ref{board1}.}
    \label{board2}
\end{figure}
\noindent The resulting magnetic field plot (where we are considering $[\cos \omega t + N(t)]$ as a re-scaled expression for the field) is displayed in Fig. \ref{sig}. It was found in the study that even after including this noise, the needles were able to attain synchronization after a time of $t = 200$ seconds; see Fig. \ref{board2}.

\section{Conclusion}
We study that for a simple system described by equation (\ref{briggseqn}) with no damping and no gravitational influence, the critical value of $\lambda$ at which the motion becomes chaotic is $\lambda_c = 1$. When a finite amount of damping is introduced ($\gamma > 0$), then this critical value of $\lambda$ increases from 1. It was also found that when gravity is included, two interesting bifurcation diagrams (one for $\omega_0$ vs. $\phi$, and the other for $\lambda$ vs. $\phi$) emerge in the resulting simulation.

It was also found that coupling two such dipoles tends to give rise to various modes of synchronization, and that some of these modes of synchronization also show up in the 2-dimensional array generalization of coupled oscillators. Finally, the synchronisation in magnetic dipoles was also seen to be resistant to noise in the driving magnetic field.



\bibliography{example}

\end{document}